\documentclass[11pt]{article}
\usepackage{graphicx}

\begin{document}

\title{A Simple Grand Unified Relation between Neutrino Mixing
and Quark Mixing}
\author{{\bf S.M. Barr} and {\bf Heng-Yu Chen} \\
Department of Physics and Astronomy and \\
The Bartol Research Institute \\ University of Delaware \\
Newark, Delaware 19716} \maketitle

\begin{abstract}
It is proposed that all flavor mixing is caused by the mixing of the
three quark and lepton families with vectorlike fermions in ${\bf 5}
+ \overline{{\bf 5}}$ multiplets of $SU(5)$. This simple assumption
implies that both $V_{CKM}$ and $U_{MNS}$ are generated by a single
matrix. The entire $3 \times 3$ complex mass matrix of the neutrinos
$M_{\nu}$ is then found to have a simple expression in terms of two
complex parameters and an overall scale. Thus, all the presently
unknown neutrino parameters are predicted. The best fits are for
$\theta_{atm} \stackrel{<}{\sim} 40^{\circ}$. The leptonic Dirac CP
phase is found to be somewhat greater than $\pi$.
\end{abstract}

It is a striking fact that the leptonic mixing angles of the MNS
matrix \cite{MNS} are much larger than the corresponding quark
mixing angles of the CKM matrix \cite{CKM}. Grand unification
suggests a simple explanation for this. In $SU(5)$, a family of
quarks and leptons is contained in the multiplets ${\bf 10} +
\overline{{\bf 5}}$, with the left-handed leptons contained in the
$\overline{{\bf 5}}$ and the left-handed quarks contained in the
${\bf 10}$. Thus, if there is more mixing among the $\overline{{\bf
5}}$ multiplets of different families than among the ${\bf 10}$
multiplets, the disparity between leptonic and quark mixing angles
would be explained. This idea can be implemented in models based on
any grand unified group, since all such groups contain $SU(5)$ as a
subgroup. Several ways of implementing this basic idea have been
proposed in the literature \cite{BabuBarr96,lopsided}.

Here we propose a model in which the three ${\bf 10} +
\overline{{\bf 5}}$ families of fermions are supplemented by three
${\bf 5} + \overline{{\bf 5}}$ pairs. (The possible existence of
such additional ``vectorlike" fermions has been much discussed in
the literature in a variety of contexts
\cite{BabuBarr96,Georgi,NelsonBarr,OtherVectorlike,Dermisek}.) The
central idea of the model proposed here is that {\it all}
inter-family mixing is caused by the mixing between the
$\overline{{\bf 5}}$ multiplets of the ordinary families and the
$\overline{{\bf 5}}$ multiplets of the additional vectorlike pairs.
As a consequence of having this common source, both quark mixing and
lepton mixing are controlled in this model by a single matrix, which
we call $A$. This matrix can be determined from the masses and
mixing angles of the quarks alone, and this allows the entire $3
\times 3$ complex mass matrix $M_{\nu}$ of the known neutrinos
(which contains 9 real physical observables) to be predicted in
terms of just two complex parameters and an overall mass scale. The
resulting formula turns out to be quite simple. In the ``flavor
basis" of the neutrinos, i.e. the basis $(\nu_e, \nu_{\mu},
\nu_{\tau})$, the neutrino mass matrix is given by

\begin{equation}
M_{\nu} \cong
\left[ \begin{array}{ccc} 1 & 0 & 0 \\
\frac{m_s}{m_d} | \overline{V}_{us}| & 1 & 0 \\
\frac{m_b}{m_d} |V_{ub}| e^{i \delta} & \frac{m_b}{m_s} |V_{cb}| & 1
\end{array} \right]
\left[ \begin{array}{ccc} q e^{i \beta} & 0 & 0
\\ 0 & p e^{i \alpha} & 0 \\
0 & 0 & 1 \end{array} \right] \left[
\begin{array}{ccc}
1 & \frac{m_s}{m_d} |\overline{V}_{us}| &
\frac{m_b}{m_d} |V_{ub}| e^{i \delta} \\
0 & 1 & \frac{m_b}{m_s} |V_{cb}| \\
0 & 0 & 1 \end{array} \right] \mu_{\nu},
\end{equation}

\noindent where $pe^{i \alpha}$, $q e^{i \beta}$, and the overall
scale $\mu_{\nu}$ are free parameters of the model, $\delta$ is the
Kobayashi-Maskawa CP phase, and the $V_{ij}$ are the CKM matrix
elements. The expression $\overline{V}_{us}$ stands for $\sin
\theta_{us} \cos \theta_{us} = V_{us} \sqrt{1 - |V_{us}|^2}$. (Since
the other quark mixing angles are small, their cosines can be set to
one, and we can use simply $V_{cb}$ and $V_{ub}$.) Note that the CKM
elements $V_{ij}$ describing quark mixing enter this formula in such
a way that if they vanish the neutrino mass matrix becomes diagonal
and the neutrino mixing vanishes also. But while these CKM mixing
parameters are small, they are here multiplied by large ratios of
quark masses, explaining naturally why the neutrino mixing is of
order one. ($\frac{m_s}{m_d} | \overline{V}_{us}| \sim 4$,
$\frac{m_b}{m_s} |V_{cb}| \sim 2$, and $\frac{m_b}{m_d} |V_{ub}|
\sim 3$.) We shall see later how this arises.

In this model, there are three families of fermions denoted by ${\bf
10}_i + \overline{{\bf 5}}_i$ ($i=1,2,3$), and three vectorlike
pairs of fermion multiplets denoted by ${\bf 5}'_A + \overline{{\bf
5}}'_A$ ($A = 1,2,3$). In the absence of the vectorlike pairs, the
Yukawa couplings and mass matrices of the three families would be
flavor diagonal, due to discrete symmetries that distinguish the
three families from each other. All flavor mixing is indirectly
caused by mass terms that mix the $\overline{{\bf 5}}_i$ with the
$\overline{{\bf 5}}'_A$. The model is defined by the following quark
and lepton Yukawa terms:

\begin{equation}
\begin{array}{ccl}
{\cal L}_{Yuk} & = & Y_i ({\bf 10}_i {\bf 10}_i) \langle {\bf 5}_H
\rangle + y_i ({\bf 10}_i \overline{{\bf 5}}_i) \langle {\bf
5}^{\dag}_H \rangle
\\ & & \\
& + & \tilde{Y}_i ({\bf 10}_i {\bf 10}_i) \langle {\bf 45}_H \rangle
+ \tilde{y}_i ({\bf 10}_i \overline{{\bf 5}}_i) \langle {\bf
45}^{\dag}_H \rangle
\\ & & \\
& + & \frac{\lambda_i}{M_R}(\overline{{\bf 5}}_i \overline{{\bf
5}}_i) \langle {\bf 5}_H \rangle \langle {\bf 5}_H \rangle
\\ & & \\
& + &  Y'_{AB} ({\bf 5}'_A \overline{{\bf 5}}'_B) \langle {\bf 1}_H
\rangle + y'_{Ai} ({\bf 5}'_A \overline{{\bf 5}}_i) \langle {\bf
1}'_{Hi} \rangle,
\end{array}
\end{equation}

\noindent where the subscript $H$ denotes Higgs multiplets. The
renormalizable Yukawa terms in Eq. (2) are the most general allowed
by the symmetry $K_1 \times K_2 \times K_3 \times K'$, where (for a
given $i$ equal to 1,2, or 3) $K_i$ is a $Z_2$ symmetry under which
${\bf 10}_i$, $\overline{{\bf 5}}_i$, and ${\bf 1}'_{Hi}$ are odd
and all other fields even. $K'$ is a $Z_N$ symmetry ($N>2$) under
which ${\bf 5}'_A \rightarrow e^{2\pi i/N} {\bf 5}'_A$,
$\overline{{\bf 5}}'_A \rightarrow e^{2 \pi i/N} \overline{{\bf
5}}'_A$, ${\bf 1}_H \rightarrow e^{-4\pi i/N} {\bf 1}_H$, and ${\bf
1}'_{Hi} \rightarrow e^{-2 \pi i/N} {\bf 1}'_{Hi}$. (This is just
one example. Many other simple discrete symmetries would give the
form in Eq. (2).)

The first four terms in Eq. (2) are the standard Yukawa terms of
$SU(5)$ grand unification, and are the minimal terms needed to give
mass to the known quarks and leptons. (As already noted in the
original Georgi-Glashow paper on $SU(5)$ unification, the presence
of a ${\bf 45}$ of Higgs fields avoids the unrealistic relations
between down quark and charged lepton masses that would arise if
only a ${\bf 5}$ of Higgs fields existed \cite{GeorgiGlashow}.) The
fifth term is the standard dimension-5 Weinberg operator that gives
the left-handed neutrinos Majorana masses \cite{Weinberg}. (The
symmetry $K'$ prevents other dimension-5 operators that would give
neutrino masses, such as $\overline{{\bf 5}}'_A \overline{{\bf
5}}'_B {\bf 5}_H {\bf 5}_H$.) Note that all these standard terms are
forced to be flavor diagonal by the $K_i$ symmetries.

The last two terms in Eq. (2) are the only ones peculiar to this
model. The first of these simply gives masses to the vectorlike
fermions, and the second gives masses that mix these vectorlike
fermions with the three families. The Higgs fields in these two
terms are gauge singlets, so that their VEVs would naturally be
superlarge. All that matters for the purposes of this paper is that
the masses coming from these two terms be roughly of the same scale,
which we shall call ${\cal M}$, and that this scale be large
compared to the masses of the down quarks and charged leptons. Note
that the Yukawa matrices in these two terms are in general {\it not}
diagonal.

This is the model; all that remains is to extract its predictions.
First, let us examine the mass matrix of the down quarks that
emerges from Eq. (2). There are left-handed anti-down quarks in both
$\overline{{\bf 5}}_i$ and $\overline{{\bf 5}}'_A$, which will be
denoted $d^c_i$ and $D^{c \prime}_A$ respectively. There are
left-handed down quarks in both ${\bf 10}_i$ and ${\bf 5}_A$, which
will be denoted $d_i$ and $D'_A$ respectively. Altogether, then,
there is a $6 \times 6$ mass matrix for the down quarks, given by

\begin{equation}
{\cal L}_{(d\; mass)} = \left( d_i \;\ D'_A \right) \; \left(
\begin{array}{cc} (m_D)_i \delta_{ij} & 0 \\
\Delta_{Aj} & M_{AB} \end{array} \right) \; \left(
\begin{array}{c} d^c_j \\ D^{c\prime}_B \end{array} \right),
\end{equation}

\noindent where $(m_D)_i = y_i \langle {\bf 5}^{\dag}_H \rangle +
\tilde{y}_i \langle {\bf 45}^{\dag}_H \rangle$, $M_{AB} = Y'_{AB}
\langle {\bf 1}_H \rangle$, and $\Delta_{Aj} = y'_{Aj} \langle {\bf
1}'_{Hj} \rangle$. Here and throughout the paper, Dirac mass
matrices are multiplied from the left by the left-handed fermions
and from the right by the right-handed fermions (or, equivalently,
the left-handed anti-fermions)

The $6 \times 6$ matrix in Eq. (3) can be block-diagonalized by
multiplying it from the right by a unitary matrix ${\cal U}$ whose
elements are of order one (since the elements of the matrices
$\Delta$ and $M$ are of the same order) and from the left by a
unitary matrix whose angles are of order $m_D/{\cal M} \ll 1$ and
which therefore can be neglected. Specifically, the unitary matrix
${\cal U}$ is such that

\begin{equation}
\left( \begin{array}{cc} m_D & 0 \\ \Delta & M \end{array} \right)
\; {\cal U} \equiv \left( \begin{array}{cc} m_D & 0 \\ \Delta & M
\end{array} \right)
\; \left( \begin{array}{cc} A & B \\
C & D
\end{array} \right) = \left( \begin{array}{cc} m_D A & m_D B \\ 0 & \tilde{M}
\end{array}  \right),
\end{equation}

\noindent where $A = [I + \Delta^{\dag} M^{-1 \; \dag} M^{-1}
\Delta]^{-1/2}$. The off-diagonal block $m_D B$ in the last matrix
in Eq. (4) can be removed by a rotation from the left that is of
order $m_D/\tilde{M}$, which is negligible, as already noted. Thus,
after block-diagonalization, the upper-left $3 \times 3$ block that
describes the masses of the three observed down quarks becomes
simply

\begin{equation}
M_D = m_D \; A.
\end{equation}

\noindent In other words, the net effect of the mixing of the three
families with the heavy vectorlike fermions is to multiply the
diagonal mass matrix $m_D$ from the right by a non-diagonal matrix
$A$. This can be understood diagramatically from Fig. 1.

\begin{figure}[h]
\begin{center}
\includegraphics[width=10cm]{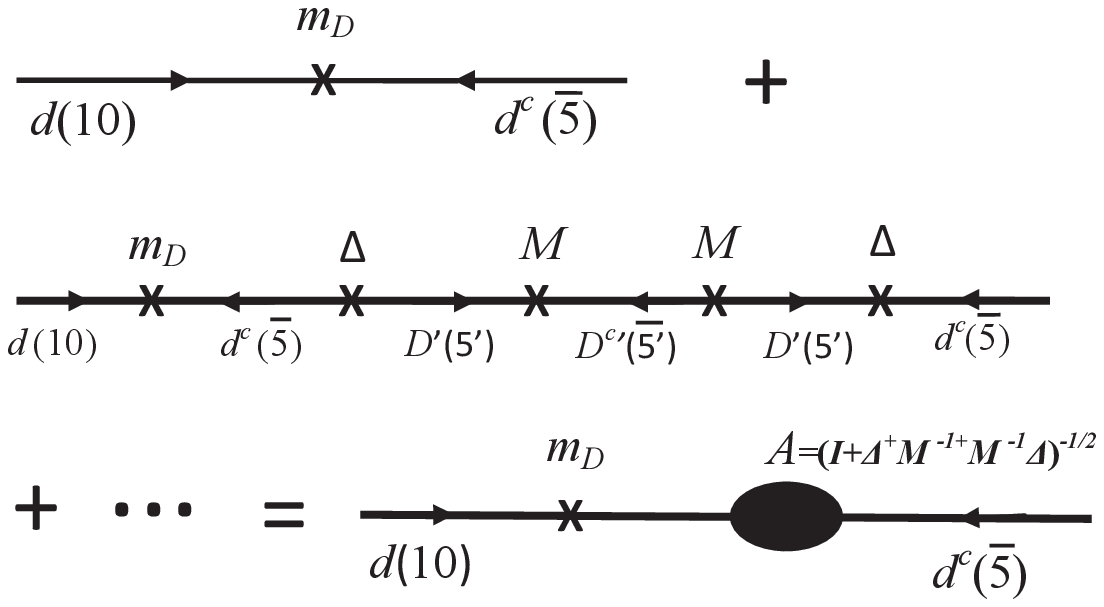}
\end{center}
\end{figure}

\vspace{1cm}

\noindent {\bf Fig. 1} Diagrams showing how the mass terms
($\Delta$) that mix the $\overline{{\bf 5}}_i$ with the
$\overline{{\bf 5}}'_A$ lead to insertions of the matrix $A$ on
external $\overline{{\bf 5}}$ fermion lines.

\vspace{0.2cm}

\noindent From these diagrams, it is easy to see that such factors
of $A$ accompany external fermion lines that are in the
$\overline{{\bf 5}}$ representation of $SU(5)$. Thus the mass
matrices for the up quarks, down quarks, charged leptons and
neutrinos (which come, respectively, from $({\bf 10} \; {\bf 10})
{\bf 5}_H$, $({\bf 10} \; \overline{{\bf 5}}) {\bf 5}_H^{\dag}$,
$(\overline{{\bf 5}} \; {\bf 10}) {\bf 5}_H^{\dag}$, and
$(\overline{{\bf 5}} \; \overline{{\bf 5}}) {\bf 5}_H {\bf 5}_H$
terms) have the form

\begin{equation}
\begin{array}{l}
M_U = m_U, \\ \\
M_D = m_D A, \\ \\
M_{\ell} = A^T m_{\ell}, \\ \\ M_{\nu} = A^T \; m_{\nu} A
\end{array}
\end{equation}

\noindent where the matrices $m_U$, $m_D$, $m_{\ell}$, and $m_{\nu}$
are all diagonal. This can also easily be shown by
block-diagonalizing the full mass matrices of the charged leptons
and neutrinos in the same way that we did for the down quarks.

It might seem that the matrix we have called $A$ should be different
for the different types of fermions due to renormalization effects.
At the unification scale, the same $3 \times 3$ matrices $\Delta$
and $M$ appear in the $6 \times 6$ mass matrices of the charged
leptons and the down quarks. But, due to gluon loops, the $\Delta$
and $M$ of the down quarks should run more strongly between the
unification scale and the scale ${\cal M}$ than the corresponding
matrices of the leptons. The crucial point, however, is that $A$
depends on the ratio $M^{-1} \Delta$; and since gauge boson loops
cause $\Delta$ and $M$ to run in the same way, these effects cancel
out in $A$. Moreover, the renormalization effects due to Yukawa
couplings (which are small for the $\overline{{\bf 5}}$ fermions)
can be neglected. Thus, it really is the same matrix $A$ that
appears in $M_D$, $M_{\ell}$, and $M_{\nu}$. Ultimately, this is due
to $SU(5)$ symmetry. It is this fact that makes this model so
predictive.

To extract the predictions of this model, let us consider first the
quarks. The up quarks are already in the ``mass basis", since $M_U =
m_U$ is diagonal. The down quark mass matrix can be written

\begin{equation}
\begin{array}{l}
M_D = m_D A = \mu_d \left( \begin{array}{ccc}
\delta_d & 0 & 0 \\
0 & \epsilon_d & 0 \\ 0 & 0 & 1 \end{array} \right) \;
\left( \begin{array}{ccc} A_{11} & A_{12} & A_{13} \\
A_{21} & A_{22} & A_{23} \\ A_{31} & A_{32} & A_{33}
\end{array}
\right) \\ \\
\longrightarrow \overline{\mu}_d \left(
\begin{array}{ccc}
\overline{\delta}_d & 0 & 0 \\
0 & \overline{\epsilon}_d & 0
\\ 0 & 0 & 1 \end{array} \right)
\; \left( \begin{array}{ccc} 1 & b & c e^{i \theta} \\
0 & 1 & a
\\ 0 & 0 & 1
\end{array} \right) = \overline{\mu}_d
\left( \begin{array}{ccc} \overline{\delta}_d & \overline{\delta}_d
b & \overline{\delta}_d c e^{i \theta} \\ 0 & \overline{\epsilon}_d
& \overline{\epsilon}_d a
\\ 0 & 0 & 1 \end{array} \right),
\end{array}
\end{equation}

\noindent where $\mu_d$ is the 33 element of the diagonal matrix
$m_D$. As we have indicated in the second line of Eq. (7), the
matrix $A$ can be brought to triangular form by rotations of the
right-handed quarks. Then, by rescaling the parameters $\mu_d$,
$\epsilon_d$, and $\delta_d$, one can make the diagonal elements of
$A$ equal to 1. And finally, by rephasings of the left-handed and
right-handed quarks, one can remove all phases except one, which we
can place in the 13 element, as shown.

Since the up quark mass matrix is already diagonal, the CKM mixing
matrix of the quarks comes entirely from diagonalizing the mass
matrix in Eq. (7). It is easy to show from that equation that
$|V_{cb}| = \sin \theta_{cb} \cong \tan \theta_{cb} \cong
\overline{\epsilon}_d a$, $|V_{ub}| = \sin \theta_{ub} \cong \tan
\theta_{ub} \cong \overline{\delta}_d c$, $|V_{us}| \sqrt{1 -
|V_{us}|^2} = \tan \theta_{us} \cong \overline{\delta}_d
b/\overline{\epsilon}_d $, and $\theta = \delta$, the CP phase of
the quarks. (Notice that the basic structure predicted by the model,
given in Eq.(6), explains why $|V_{ub}| \sim |V_{us} V_{cb}|$, since
$|V_{us}| = O(\overline{\delta_d}/\overline{\epsilon_d})$, $|V_{cb}|
= O(\overline{\epsilon_d})$, and $|V_{ub}| =
O(\overline{\delta_d})$.)

It is also clear from Eq. (7) that $\overline{\epsilon}_d \cong
m_s/m_b$ and $\overline{\delta}_d \cong m_d/m_b$. Notice that this
implies that $\overline{\delta}_d \ll \overline{\epsilon}_d \ll 1$,
so that the matrix $m_D$ is not only diagonal (because of family
symmetry) but ``hierarchical". (Though the model as presented in Eq.
(2) does not explain the mass hierarchy among the families, we shall
see that a simple extension of the model can do this.) Combining the
above relations gives

\begin{equation}
\begin{array}{l}
a \cong \frac{m_b}{m_s} |V_{cb}| \sim 2, \\
b \cong \frac{m_s}{m_d} |V_{us}| \sqrt{1 - |V_{us}|^2} \sim 4, \\
c \cong \frac{m_b}{m_d} |V_{ub}| \sim 3, \\  \theta \cong \delta.
\end{array}
\end{equation}

Turning to the leptons, one sees that the mass matrix of the charged
leptons can be written

\begin{equation}
\begin{array}{l}
M_{\ell} = A m_{\ell} = \left( \begin{array}{ccc} A_{11} & A_{21} & A_{31} \\
A_{12} & A_{22} & A_{32} \\ A_{13} & A_{23} & A_{33}
\end{array} \right) \; \left( \begin{array}{ccc}
\delta_{\ell} & 0 & 0 \\
0 & \epsilon_{\ell} & 0 \\ 0 & 0 & 1
\end{array} \right) \mu_{\ell}
\\ \\ \longrightarrow  \left( \begin{array}{ccc}
1 & 0 & 0 \\
b & 1 & 0 \\ c e^{i \theta} & a & 1
\end{array}
\right) \; \left( \begin{array}{ccc}
\overline{\delta}_{\ell} & 0 & 0 \\
0 & \overline{\epsilon}_{\ell} & 0
\\ 0 & 0 & 1 \end{array} \right) \overline{\mu}_{\ell}
=  \left( \begin{array}{ccc}
\overline{\delta}_{\ell} & 0 & 0 \\
\overline{\delta}_{\ell} b
& \overline{\epsilon}_{\ell} & 0 \\
\overline{\delta}_{\ell} c e^{i \theta} & \overline{\epsilon}_{\ell}
a & 1
\end{array} \right) \overline{\mu}_{\ell}.
\end{array}
\end{equation}

\noindent We have transformed $A$ to have the same form as in the
second line of Eq. (7), by doing the same combination of rotations
to the left-handed leptons as we did to the right-handed down
quarks, followed by analogous rescalings and rephasings. If we do
the same rotations to the left-handed charged leptons and
left-handed neutrinos, no MNS mixing is induced at this stage; but
by doing so, it is clear that we also make the matrix $A$ appearing
in $M_{\nu} = A^T m_{\nu} A$ have the same form as in the second
line of Eq. (7) and Eq. (9). What results is precisely the form
shown in Eq. (1) for the neutrino mass matrix. Since we only have
the freedom to rephase the left-handed neutrinos, there are {\it
three} physical phases in Eq. (1), rather than one as in the other
mass matrices. The extra two phases are the ones called $\alpha$ and
$\beta$ in Eq. (1).

The hierarchy among the charged lepton masses tells us that
$\overline{\delta}_{\ell} \ll \overline{\epsilon}_{\ell} \ll 1$. So,
the diagonal matrix $m_{\ell}$ is hierarchical, just as $m_D$ and
$m_U$ are. By comparing Eqs. (7) and (9), we see how this model
explains the disparity between the neutrino mixing angles and quark
mixing angles. Because $M_D = m_D A$ , whereas $M_{\ell} = A^T
m_{\ell}$, the mass matrix of the down quarks has a hierarchy among
the {\it rows}, whereas the charged lepton mass matrix has a
hierarchy among the {\it columns}. Since rotations of the {\it
left-handed} fermions (which are the ones relevant to the CKM and
MNS mixing angles) are rotations among the {\it rows}, we see that
small quark mixing angles and large lepton mixing angles arise.
(This is a realization of the basic idea of ``lopsided" models
\cite{BabuBarr96, lopsided}.)

The charged lepton mass matrix in the second line of Eq. (9) is not
yet diagonal, but to a very good approximation it can be
diagonalized by rotations done only to the right-handed charged
leptons. Rotations of the left-handed charged leptons are also
required, but they are rotations by angles that are proportional to
$\overline{\epsilon}^2_{\ell}$,
$(\overline{\delta}_{\ell}/\overline{\epsilon}_{\ell})^2$, and
$\overline{\delta}^2_{\ell}$. The only one of these that is
numerically significant is a rotation in the $\mu_L \tau_L$ plane by
angle $\theta_{\mu \tau} \cong a \overline{ \epsilon}^2_{\ell} \sim
2 (m_{\mu}/m_{\tau})^2 \sim 0.4^{\circ}$. This contributes at the
1\% level to the atmospheric neutrino mixing angle, and this has
only a minor effect on the predictions of the model, as we shall
see.

Thus, in effect, the mass matrix in the second line of Eq. (9) is in
the mass basis of the left-handed charged leptons. Consequently, the
mass matrix $M_{\nu}$ shown in Eq. (1) contains all the information
about the masses, mixings and CP-violating phases of the neutrinos.
There are nine physical observables involved: the three neutrino
masses, the three MNS angles, the Dirac CP phase, and the two
majorana CP phases of the neutrinos. These are all determined by the
five model parameters in Eq. (1): $\mu_{\nu}$, $p e^{i \alpha}$, and
$ q e^{i \beta}$.

Since five neutrino observables have already been measured
($\theta_{sol}$, $\theta_{atm}$, $\theta_{13}$, $\delta m^2_{23}$
and $\delta m^2_{12}$), we can use them to determine the five model
parameters, and then predict the four as-yet-unmeasured neutrino
observables. Since the equations are non-linear (they involve
trigonometric functions), there is no guarantee that they can fit
the five measured neutrino properties with five adjustable
parameters. (To put it another way, the adjustable parameters
$\alpha$ and $\beta$ are bound within the range $[0, 2 \pi)$.)
Nevertheless, good fits are obtained. This is only true, however, if
some of the measured neutrino properties have values that lie within
a smaller range than that presently allowed by experiment. For
example, although the current experimental range of the atmospheric
neutrino mixing angle is $\theta_{atm} = 45 \pm 6.5^{\circ}$
\cite{pdg}, the model only obtains good fits for $\theta_{atm}
\stackrel{<}{\sim} 43^{\circ}$, with values near $40^{\circ}$
preferred, as we shall see. The fits also prefer a value of
$\theta_{sol}$ greater than or equal to $34^{\circ}$, i.e. greater
than the present experimental central value. The quark properties
are also constrained: the best fits are obtained with $m_s/m_d
\stackrel{<}{\sim} 20$, and $\delta$ greater than or equal to its
present experimental central value. Thus, in addition to predicting
the four as-yet-unmeasured neutrino observables, the model places
non-trivial and testable constraints on the values of quantities
that have been measured.

In Table I, we show a representative fit in which all the input
quark parameters and the neutrino observables obtained as output are
in their experimentally allowed ranges (and in most cases at their
central values). The experimental values are taken from the 2012
Review of Particle Properties \cite{pdg}, except for $\delta_{lep}$
(the neutrino Dirac CP phase) where we use the result of a recent
global analysis of neutrino data \cite{GlobalAnalysis}. For
$m_b/m_s$ we have used the renormalization group results of
\cite{RGE} to obtain $m_s(m_b)$ from $m_s(2 {\rm GeV})$, which is
given in \cite{pdg}.

\noindent {\bf Table I.} A fit to the quark and neutrino data.
$\mu_{\nu}$, $p e^{i \alpha}$, and $q e^{i \beta}$ are model
parameters. $\delta_{lep}$ is the neutrino Dirac CP phase, and
$(M_{\nu})_{ee}$ the mass that comes into neutrinoless double beta
decay.

\[
\begin{tabular}{|l|l|l|}
\hline {\bf Quantity} & {\bf Values in fit} & {\bf Experiment} \\
\hline \hline $\mu_{\nu}$ & 0.1428 eV & --- \\
\hline $p e^{i \alpha}$ & $0.1525e^{-2.734 i}$ & --- \\
\hline $q e^{i \beta}$ & $0.01405 e^{-0.352 i}$ & --- \\
\hline \hline $m_b/m_s$ & 52.9 & $52.9 \pm 2.6$ \\
\hline $m_s/m_d$ & 19 & 17 {\rm to} 22\\
\hline $|V_{us}|$ & 0.2252 & $0.2252 \pm 0.0009$ \\
\hline $|V_{cb}|$ & 0.0409 & $0.0409 \pm 0.0011$ \\
\hline $|V_{ub}|$ & 0.00415 & $0.00415 \pm 0.00049$ \\
\hline $\delta$ & 1.30 rad & $1.187^{+0.175}_{-0.192}$ rad \\
\hline \hline $\theta_{sol}$ & $34.1^{\circ}$ & $33.89^{\circ}
\;^{+0.976^{\circ}}_{-0.971^{\circ}}$ \\
\hline $\theta_{atm}$ & $40^{\circ}$ & $45^{\circ} \pm 6.5^{\circ}$  \\
\hline $\theta_{13}$ & $9.12^{\circ}$ & $9.122^{\circ}
\;^{+0.609^{\circ}}_{-0.647^{\circ}}$  \\
\hline $\delta m^2_{23}$ & $2.32 \times 10^{-3}$ eV$^2$ &
$2.32^{+0.12}_{-0.08} \times 10^{-3}$ eV$^2$ \\
\hline $\delta m^2_{12}$ & $7.603 \times 10^{-5}$ eV$^2$ &
$(7.5 \pm 0.2) \times 10^{-5}$ eV$^2$ \\
\hline \hline $\delta_{lep}$ & $1.15 \pi$ rad & $1.1 \pi^{+0.3 \pi}_{-0.4 \pi}$ rad \\
\hline $(M_{\nu})_{ee}$ & 0.0020 eV & \\
\hline
\end{tabular}
\]

\noindent Note that the model's prediction for $\delta_{lep}$ is
$1.15 \pi$ radians, which accords remarkably well with the one-sigma
range found in \cite{GlobalAnalysis} of $1.1 \pi^{+0.3 \pi}_{-0.4
\pi}$ rad. The value of $(M_{\nu})_{ee}$ (to which the amplitude of
neutrinoless double beta decay is proportional) is much smaller than
the experimental limits, which tend to be in the range of a few
tenths of an eV to several eVs for different experiments \cite{pdg}.
This prediction of the model is not very sensitive to variation of
the model's input parameters.

Figs. 2-4 show the degree of sensitivity of the $\delta_{lep}$
prediction to the values of $\theta_{atm}$, $\theta_{sol}$, and
$\delta$ (the quark CP phase). In Fig. 2, we have fixed the values
of all the quark mass ratios and CKM parameters, and of
$\theta_{sol}$ and $\theta_{13}$, but have allowed $\theta_{atm}$
and the ratio $\delta m^2_{12}/\delta m^2_{23}$ (which we henceforth
call $r$) to take different values. The curves are the relation of
$r$ to the predicted $\delta_{lep}$ for different values of
$\theta_{atm}$. The horizontal lines are the one-sigma limits for
$r$. One sees that $\delta_{lep}$ is predicted to be roughly $1.15
\pi$ radians and that values of $\theta_{atm} \stackrel{<}{\sim}
41^{\circ}$ are preferred.

In Fig. 3, we have done a similar thing, but this time fixing
$\theta_{atm}$ to be $40^{\circ}$ and allowing $\theta_{sol}$ and
$r$ to vary. One can see a preference for values of $\theta_{sol}$
equal or above the present experimental central value. In Fig. 4, we
have allowed the quark CP phase $\delta$ and $r$ to vary. One sees
that the best-fit value of $\delta_{lep}$ is rather insensitive to
the assumed values of the measured quark and neutrino properties,
but the width of the range of $\delta_{lep}$ values that give good
fits is quite sensitive. A more detailed analysis of the predictions
of the model will be given in another paper.

\begin{figure}[h]
\begin{center}
\includegraphics[width=8cm]{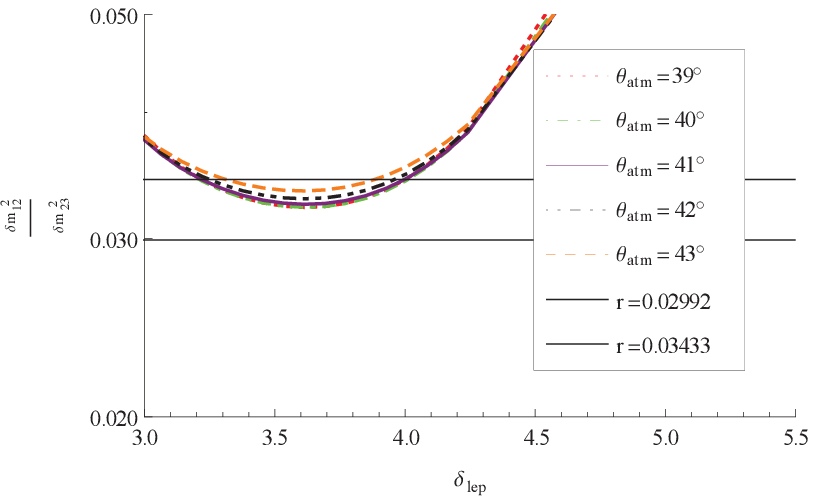}
\end{center}
\end{figure}

\noindent {\bf Fig. 2} The result of fits with the values of quark
parameters given in Table I, $\theta_{sol} = 34.2^{\circ}$, and
$\theta_{13} = 9.12^{\circ}$. The curves are the relation of $r$
($\equiv \delta m^2_{12}/\delta m^2_{23}$) to the predicted
$\delta_{lep}$ for different values of $\theta_{atm}$. The
horizontal lines are the one-sigma limits for $r$.

\begin{figure}[h]
\begin{center}
\includegraphics[width=8cm]{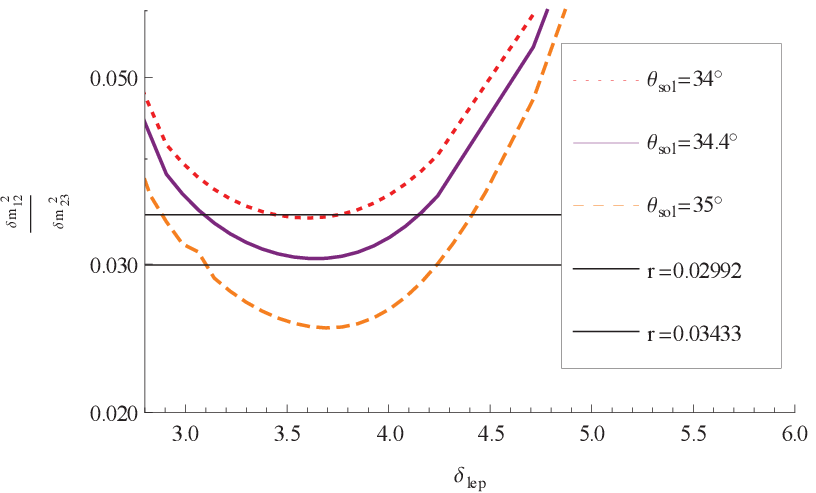}
\end{center}
\end{figure}

\noindent {\bf Fig. 3} The result of fits with the values of quark
parameters given in Table I, $\theta_{atm} = 40^{\circ}$, and
$\theta_{13} = 9.12^{\circ}$. The curves are the relation of $r$
($\equiv \delta m^2_{12}/\delta m^2_{23}$) to the predicted
$\delta_{lep}$ for different values of $\theta_{sol}$.

\newpage

\begin{figure}[h]
\begin{center}
\includegraphics[width=8cm]{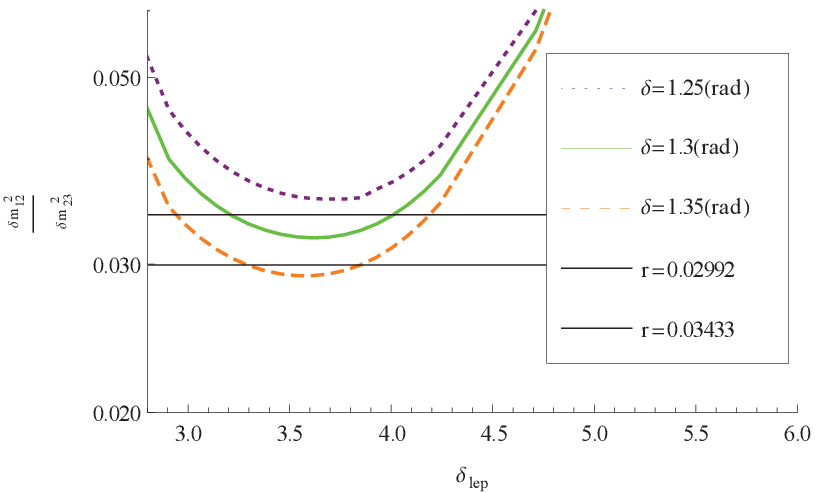}
\end{center}
\end{figure}

\noindent {\bf Fig. 4} The result of fits with the values of quark
parameters other than $\delta$ given in Table I, $\theta_{sol} =
34.2^{\circ}$, and $\theta_{13} = 9.12^{\circ}$, and $\theta_{atm} =
40^{\circ}$. The curves are the relation of $r$ ($\equiv \delta
m^2_{12}/\delta m^2_{23}$) to the predicted $\delta_{lep}$ for
different values of the quark phase $\delta$.

\vspace{0.2cm}

Besides its great simplicity, one feature of the model proposed here
that increases its plausibility is that it allows the simple
solution to the Strong CP Problem proposed in \cite{NelsonBarr}. All
that is required is that CP be assumed to be a symmetry of the
Lagrangian that is spontaneously broken by the VEVS of the singlet
scalars ${\bf 1}'_{Hi}$ that produce the off-diagonal mass matrix
$\Delta$ in Eq. (3). In fact, the model proposed here has the same
structure as the model originally proposed by A.E. Nelson in
\cite{NelsonBarr}, except that here the $3 \times 3$ mass matrices
of the ordinary three families are required to be flavor diagonal by
family symmetries.

The model proposed here gives an account of how the CKM and MNS
flavor mixings arise, but does not explain the mass hierarchy among
the families, since the hierarchies in the diagonal matrices $m_U$,
$m_D$, $m_{\ell}$, and $m_{\nu}$ are simply assumed. There are,
however, several simple ways in which the present model could be
extended to give an explanation of the mass hierarchy. One way is to
combine the structure in this model with the structure assumed in
\cite{BabuBarr96}. In that paper, the mass hierarchies were
explained by the three ordinary families mixing with vectorlike
${\bf 10} + \overline{{\bf 10}}$ fermion pairs in a way analogous to
the mixing with ${\bf 5} + \overline{{\bf 5}}$ assumed here.
Combining the structures of the two models would be appealing since
it would mean that the vectorlike fermions would comprise entire
family-antifamily pairs. (It has been pointed out that this can lead
in a simple way to unification of gauge couplings in non-SUSY models
\cite{Dermisek}.)

Another possibility would be a Froggatt-Nielsen scheme
\cite{FroggattNielsen}. For example, instead of the family symmetry
assumed above, suppose that the $K_i$ ($i = 1$ or 2) were $Z_4$
symmetries, under which $\overline{{\bf 5}}_i \rightarrow -
\overline{{\bf 5}}_i$, ${\bf 1}'_{Hi} \rightarrow - {\bf 1}'_{Hi}$,
${\bf 10}_i \rightarrow i {\bf 10}_i$, and $S_i \rightarrow i S_i$,
where the $S_i$ are Froggatt-Nielsen fields that are $SU(5)$
singlets. Effective Yukawa terms containing factors of ${\bf 10}_i$
would then have to contain equal numbers of factors of $\langle S_i
\rangle/M_F \equiv \epsilon_i$, where $M_F$ is some flavor-physics
scale. If $\epsilon_1 \ll \epsilon_2 \ll 1$, a mass hierarchy among
families would result. Moreover, the hierarchy would be strongest
for the up quarks (for which it is quadratic in the $\epsilon_i$),
intermediate for the down quarks and charged leptons (for which it
is linear in the $\epsilon_i$), and weakest for the neutrinos (which
involve no factors of $\epsilon_i$).  This is just the pattern that
is observed. One should note that the same relationship among the
mass hierarchies is obtained in the approach of \cite{BabuBarr96}.

Finally it should be noted that the present model could be embedded
in many grand unified schemes. For example, in an $SO(10)$ model,
the ordinary families could be in three ${\bf 16}$ multiplets, while
the vectorlike fermions could be in three ${\bf 10}$ multiplets. In
$E_6$, one gets the extra vectorlike fermions ``for free", since the
${\bf 27}$ contains ${\bf 16} + {\bf 10} + {\bf 1}$ of $SO(10)$.
Different patterns of breaking of the grand unified group could be
assumed without affecting the predictions for fermion masses and
mixings. For example, in many unified models, an adjoint Higgs field
does some of the breaking of the unification group. If that adjoint
Higgs multiplet does not transform under the $K'$ symmetry mentioned
after Eq. (2), it would not couple renormalizably to $({\bf 5}'_A
\overline{{\bf 5}}'_B)$ or $({\bf 5}'_A \overline{{\bf 5}}_i)$ and
hence not contribute to the matrices $\Delta$ and $M$ in Eq. (4) and
the matrix $A$. Consequently, except for negligible higher-order
corrections, the matrix $A$ would not ``know" that the unification
group is broken, and the same $A$ would appear in both the quark and
lepton sectors, as is necessary for the model to be predictive.

In conclusion, if all flavor changing in both the quark and lepton
sectors arises as a consequence of the mixing of the ordinary
families with vectorlike fermions that are in ${\bf 5} +
\overline{{\bf 5}}$ of $SU(5)$, a testable relationship arises
between the quark and lepton mixing. This relationship allows the
prediction of the four as-yet-unmeasured neutrino observables as
well as testable constraints on several quantities that have been
measured. Measurement of the Dirac CP phase of the neutrinos
$\delta_{lep}$, as well as more precise determinations of such
quantities as $\theta_{atm}$, $\theta_{sol}$, $|V_{ub}|$, $m_s/m_d$,
and $\delta$ (the quark CP phase) would provide stringent tests of
the model.

\section*{Acknowledgements} The authors acknowledge useful discussions
with the participants of the CETUP2012 workshop. This work was
supported by U.S. DOE under contract DE-FG02-12ER41808.

\end{document}